\documentstyle[12pt,epsf]{article}

\def\m{\mu}
\def\n{\nu}
\def\r{\rho}

\def\p{\phi}
\def\vp{\varphi}
\def\th{\theta}
\def\b{\beta}
\def\a{\alpha}
\def\l{\lambda}
\def\pa{\partial}
\def\ep{\epsilon}

\def\be{\begin{eqnarray}}
\def\ee{\end{eqnarray}}
\def\nn{\nonumber}

\def\ll{\left}
\def\rr{\right}
\def\fr{\frac}

\def\ci{\cite}
\def\bi{\bibitem}

\begin{document}

\title
{{\hskip 9.5cm {\normalsize La Plata-Th 99/11}} \\ 
{\vskip -.3cm \hskip 10cm  {\normalsize IP-BBSR 99/32}} \\
{\bf \LARGE Monopoles in non-Abelian Einstein-Born-Infeld Theory}}
\author{Prasanta~K.~Tripathy\thanks{email: prasanta@iopb.res.in} \\
\normalsize{\it Institute of Physics, Bhubaneswar 751 005, India} \\
and\\
Fidel~A.~Schaposnik\thanks{email: fidel@athos.fisica.unlp.edu.ar}  \\
\normalsize{\it 
Departamento de F\'\i sica, Universidad Nacional de La Plata} \\ 
\normalsize{\it C.C. 67, 1900 La Plata, Argentina}} 

\maketitle

\begin{abstract}
We study   static spherically symmetric monopole solutions in 
non-Abelian Einstein-Born-Infeld-Higgs model with  normal trace
structure. These monopo\-les are similar to the corresponding solution
with symmetrised trace structure and are existing only up to some 
critical value of the strength of the gravitational interaction.
In addition, similar to their flat space counterpart, they also admit 
a critical value of the Born-Infeld parameter $\b $.
\end{abstract}


The Dirac-Born-Infeld (DBI) model\ci{born,infeld} has recently received special 
consideration 
in connection with string theory and D-brane dynamics\ci{dai,joep} after
the discovery that the low energy effective action
for D-branes is precisely described by the DBI action 
\cite{fradkin}-\cite{leigh}
(for a review see \ci{aat} ).  BPS solutions
to DBI theory for gauge fields and scalars
were then studied  and interpreted in terms of branes
pulled by strings \cite{juan}-\cite{Hashi}.

Different non-Abelian generalization of the DBI action have been proposed
\cite{nappi}-\cite{park}. They differ in the way a scalar action is defined
from objects carrying group indices. The use of a symmetrised trace proposed
by Tesytlin \cite{aat2} in the context of superstring theory seems to be the
natural one in connection with supersymmetry and  leads to a linearized 
Lagrangian with BPS equations identical to Yang-Mills
ones \cite{brecher}-\cite{chr}. The use of the ordinary trace and other
recipes for defining the DBI action have been also investigated
\cite{park}.

Vortex, monopole and other soliton-like solutions to 
theories containing a DBI action have been studied  for Abelian
and non-Abelian gauge symmetry \cite{nakamura}-\cite{galtsov}, 
in this last case
both using the symmetrised and the normal trace operation to define
a scalar DBI action. A distinctive feature was discovered for
DBI vortices and monopoles: 
there exists  a critical value $\beta_c$ of the  
Born-Infeld $\beta$-parameter below which regular solutions cease
to exist \cite{moreno}-\cite{pakman}. In the non-Abelian case, this critical
value can be  tested only when the normal trace is adopted since, for the
symmetrised trace, the Lagrangian 
is known only as a perturbative series in $1/\beta$ and   the results 
are valid  outside the domain where $\beta$-criticality takes
place.

The existence of $\beta_c$  is much resemblant to the phenomenon occuring
for self-gravitating monopoles \cite{ortiz}-\cite{volkov}, 
that exhibit a critical behavior
with respect to the strength of the gravitational 
interaction: above some maximum value of a parameter $\a$, 
related to the strength of gravitational 
interaction, regular monopole
solutions collapse. This behavior,
 originally described in Einstein-Yang-Mills-Higgs
theory,  has been shown to take place also when the DBI
action governs the dynamics of the gauge field \cite{pkt1,pkt2}.
Since the symmetric trace has been adopted in these last references, 
the existence of $\beta_c$ in DBI theories coupled to gravity 
could  not be analysed; 
it is the purpose of the present work to
address this issue, namely, to study the Einstein-DBI-Higgs model using 
the normal trace for defining the DBI action, find self-gravitating monopole
solutions, determine whether they cease to exist below
some critical value $\beta_c$ as in the flat space
case  and, in the affirmative,  study the
interplay between $\a$ and $\beta_c$.

\vspace{0.7 cm}


We consider the following Einstein-Born-Infeld-Higgs action for $SU(2)$
fields with the Higgs field in the adjoint representation

\be
S = \int d^4x \sqrt{-g} \ll[L_G + L_{BI} + L_H \rr]
\ee
with
\be
L_G & = & \fr{1}{16\pi G}{\cal R} , \nn \\
L_H & = & -\fr{1}{2} D_{\m}\p ^a D^{\m}\p ^a
          -\fr{e^2g^2}{4}\ll(\p ^a\p ^a - v^2 \rr)^2 \nn
\ee
and the non Abelian Born-Infeld Lagrangian,
\be
L_{BI} = \b ^2 tr\ll( 1 - \sqrt{ 1
+ \fr{1}{2\b ^2}F_{\m\n}F^{\m\n}
- \fr{1}{8\b ^4}\ll(F_{\m\n}\tilde{F}^{\m\n}\rr)^2}\rr) \label{2}
\ee
where
\be
D_{\m}\p ^a = \pa _{\m}\p ^a + e \ep ^{abc} A_{\m}^b\p ^c , \nn
\ee
\be
F_{\m\n} = F_{\m\n}^a t^a
= \ll(\pa _{\m}A_{\n}^a - \pa_{\n}A_{\m}^a
+ e \ep ^{abc}A_{\m}^bA_{\n}^c\rr)t^a \nn
\ee
and the trace $tr$ in Lagrangian (\ref{2}) is defined so that $tr (t^a t^b) =
\delta^{ab}$.

Here we are interested in purely magnetic configurations, hence 
we have $F_{\m\n}\tilde{F}^{\m\n} = 0 $. The elementary excitations
are a massless photon, two massive charged vectors bosons with 
mass  $M_W = e v $, and a massive neutral Higgs scalar with 
mass $M_H = \sqrt{2} g e v $. Varying the action with respect to the 
matric we obtain the following expression for the energy-momentum 
tensor
\be
T_{\l\r} = -\fr{g^{\m\n}F_{\m\l}^{a}F_{\n\r}^{a}}{\sqrt{1
+ \fr{1}{4\b ^2}F_{\m\n}^a F^{a \m\n}}}
- 2\b ^2 g_{\l\r}\ll(1 - \sqrt{1 
+ \fr{1}{4\b ^2}F_{\m\n}^{a}F^{a \m\n}}\rr)
\ee

For static spherical symmetric solutions, the metric can be
parametrized as\ci{dieter, komar}

\be
ds^2 = -e ^{2\n(R)}dt^2 + e ^{2\l(R)}dR^2
+ r^2(R)(d\th ^2 + \sin ^2\th d\vp ^2)
\ee
We consider the 't~Hooft-Polyakov
  ansatz for the gauge  and scalar fields

\be
A_{t}^a(R) = 0 = A_{R}^a, A_{\th}^a = e_{\vp}^a\fr{W(R) - 1}{e},
A_{\vp}^a = -e_{\th}^a\fr{W(R) - 1}{e}\sin\th ,
\ee
and
\be
\p ^a = e_{R}^a v H(R) .
\ee
Putting the above ansatz in Eq.1, defining $\a ^2 = 4\pi Gv^2$
and rescaling $R \rightarrow R/ev$, $ r(R) \rightarrow r(R)/ev$ 
and $\b \rightarrow \b ev^2 $ we get the following expression for 
the Lagrangian
\be
\int dR e^{\n +\l}\ll[\fr{1}{2}\ll(1
+ e^{-2\l}\ll((r')^2 + \n '(r^2)'\rr)\rr)
- \a ^2 r^2 \ll\{ 2 \b ^2 \ll( 1 - 
\sqrt{1 + \fr{V_1}{\b ^2}}\rr) - V_2 \rr\}\rr]
\ee
where
\be
V_1 = \fr{1}{r^2} e^{-2\l }(W')^2 + \fr{1}{2r^4} ( W^2 - 1)^2 
\ee
and
\be
V_2 = \fr{1}{2} e^{-2\l }(H')^2 + \fr{1}{r^2}(HW)^2 
+ \fr{1}{4} g^2 ( H^2 - 1)^2 
\ee
Here the prime denotes differentiation with respect to $R$.
The dimensionless parameter $\a $ can be expressed as the mass ratio
\be
\a = \sqrt{4\pi}\fr{M_W}{eM_{Pl}}
\ee
where $M_{Pl} = 1/ \sqrt{G} $ is the Planck mass.
As expected in the limit of $\b \rightarrow \infty $ the above action 
reduces to that of the Einstein-Yang-Mills-Higgs model.
Also, the limit $\a = 0$, for which we must have $\n (R) = 0 = \l (R)$ 
corresponds to the flat space Born-Infeld-Higgs theory\ci{grandi}. Note
that the use of the normal trace allowed to reaccomodate
the Lagrangian in terms of a square root without reference to the
gauge group generators; this is not possible for the
case of the symmetyrized trace.

From now on we consider the gauge $r(R) = R $, corresponding to 
the Schwarzs\-child-like coordinates and rename $R = r $.
We define $A = e^{\n + \l}$ and $N = e^{-2\l}$. Integrating the 
$tt$ component of the energy-momentum we get the mass of the 
monopole equal to $M/evG$ where
\be
M = \a ^2 \int_{0}^{\infty} dr r^2
\ll\{ V_2 - 2\b ^2\ll(1 - \sqrt{1 + \fr{V_1}{\b ^2}}\rr)\rr\}
\ee

Following 't~Hooft the electromagnetic $U(1)$ field strength
${\cal F}_{\m\n}$ can be defined as
\be
{\cal F}_{\m\n} = \fr{\p ^aF_{\m\n}^a}{\mid \p \mid}
- \fr{1}{e\mid\p\mid ^3}\epsilon^{abc}\p ^aD_{\m}\p ^bD_{\n}\p ^c. \nn
\ee
Then using the ansatz(3)  the magnetic field
\be B^i = \fr{1}{2}\epsilon ^{ijk}{\cal F}_{jk} \nn \ee is equal to
$ {e_{r}^{i}}/{er^2} $ with a total flux
$4\pi /e $ and unit magnetic charge.

The $tt$ and $rr$ components of Einstein's equations are
\be
&&\fr{1}{2}\ll(1 - (rN)'\rr)  = \a ^2 r^2
\ll\{ V_2 - 2\b ^2\ll(1 - \sqrt{1 + \fr{V_1}{\b ^2}}\rr)\rr\} \\
&& \fr{A'}{A}  = \a ^2 r \ll[ (H')^2 + \fr{2 (W')^2}{r^2
\sqrt{1+\fr{V_1}{\b ^2}}}\rr]
\ee
and the matter field equations are  
\be
&&\ll(A N \fr{W'}{\sqrt{1+\fr{V_1}{\b ^2}}}\rr)'
= W A \ll( H^2 + \fr{W^2 - 1}{r^2 \sqrt{1+\fr{V_1}{\b ^2}}}\rr) \\
&&(ANr^2H')'  = A H \ll(2W^2 + g^2r^2(H^2 - 1)\rr)
\ee
Note that the field $A$ can be elliminated from the matter field 
equations using Eq.(12).


Now we consider the globally regular solution to the above equations.
Expanding the fields in powers of $r$ and keeping the leading order 
terms we obtain the following expressions near the origin
\be
&& H = a r + O(r^3), \\
&& W = 1 - b r^2 + O(r^4), \\
&& N = 1 - c r^2 + O(r^4)
\ee
where $c$ is expressed interms of the free parameters $a$ and $b$ as
\be
c = \a ^2 \ll[ a^2 + \fr{g^2}{6} 
+ \fr{4}{3}\b ^2 \ll( \sqrt{1 + \fr{6 b ^2}{\b ^2}} - 1\rr)\rr]
\ee

We are looking for asymptotically flat solution and hence we impose
\be
N = 1 - \fr{2M}{r}.
\ee
Then the gauge and the Higgs fields has the following behaviour in
the asymptotically far region:
\be
&& W = C r^{-M}e^{-r}\ll(1 + O\ll(\fr{1}{r}\rr)\rr) \\
&& H = \left\{ \begin{array}{ll}
1 - B r^{-\sqrt{2}gM - 1} e^{-\sqrt{2}gr}, & for ~~~ 0< g \le \sqrt{2} \\
1 - \fr{C^2}{g^2 - 2} r^{-2M-2} e^{-2r},   & otherwise.
\end{array}
\right.
\ee

We solved the equations of motion numerically using the above boundary 
conditions. The solutions are pretty much the same as those for the
case of the symmerised trace\ci{pkt1} for $\b \sim 1$ and they agree with the
corresponding solution for the Yang-Mills-Higgs case for lagre $\b $
\ci{peter,dieter}.
For a definite value of $\b $, the solution exists up to some critical 
value $ \a _{max} $ of the parameter $\a $. The minima of the metric 
function decreases as we increase the value of $\a $ for $\a < \a _{max} $ 
and it approches zero for $\a \rightarrow \a _{max} $. The solution 
does not exist for $ \a > \a _{max}$. It is observed that when we 
decrease the value of $\b $ the $\a _{max} $ goes on increaseing.
The values of $\a _{max}$ for different $\b $ are given in table 1.
However it is also found that, similar to the flat space monopoles\ci{grandi} 
there is a critical value $\b _c $ for finite $\a $, and the solutions
does not exist for $\b < \b _c $. For $\a = 1 $ and $g = 0 $ we find
$\b _c \sim 0.1 $ which is much smaller than the corresponding value
for flat space which is approximately $0.5$. The profile for different 
values of $\a , \b $ for the case of $g = 0 $ are given in the 
Figs.$1,2\&3$.

To understand these results, let us recall that the existence of 
an upper bound
$\a_{max}$ for the gravitation strength can be interpreted by observing
that  monopole mass$/$radius ratio increases as $\a$
increases so that $\a_{max}$ can be seen as the value at which the
monopole becomes gravitationally unstable
and  collapses \ci{ortiz}-\cite{LW}. This for Yang-Mills action, which
corresponds to the   $\beta \to \infty$ limit  of DBI action.
 Now, as $\beta$ decreases from its limiting value, 
the mass of the monopole decreases (see Table 2) so that the collapse 
should occur for a value $\a_{max}^\beta > \a_{max}^{\beta =\infty}$, 
as observed in our solutions. The same
kind of analysis can be performed regarding the lowering of $\beta_c$
as $\alpha$ increases: as shown in
\cite{moreno}-\cite{pakman}, the singular behavior occuring 
at $\beta_c$ manifests as an abrupt descent in the soliton mass, 
a phenomenon which is in concurrence with the enhancing of 
the mass as $\a$ grows. That is the reason why 
$\beta_c^\alpha < \beta_c^{\alpha=0}$.

In summary, we analysed the gravitating monopole solutions 
in non-Abelian Born-Infeld-Higgs system with a normal trace structure.
The solutions exist up to some critical value $\b _c$ of the Born-Infeld 
parameter $\b $ below which there is no solution. It was not
possible to study this feature 
in the corresponding model with symmetrised trace
structure since the perturbative expansion implicit 
in this case is not valid for small $\b $. 
We also found that the parameter $\a $ has some maximum 
allowed value $\a _{max}$ for a definite $\b $. This $\a _{max}$ 
increases as we decrease $\b $. It would be worth studying if the 
similar behaviour occures in case of non-Abelian black holes. Also 
it should be possible to study the dyon solutions in the non-Abelian
model with normal trace structure.

\vspace{2 cm}

\underline{Acknowledgements}: F.A.S is 
partially  supported by CICBA, CONICET (PIP 4330/96), ANPCYT
(PICT 97/2285). P.K.T is very much grateful to Avinash Khare for
many helpful discussions.


\vspace{.1in}

\begin{tabular}{||l|l||}
\hline\hline & \\
$\b $ & $\a _{max}^2$ \\
\hline\hline & \\
0.10  & 8.1           \\
0.15  & 7.2	      \\
0.20  & 6.5           \\
0.25  & 6.1           \\
0.30  & 5.9           \\
0.50  & 5.7           \\
1.00  & 5.6           \\
\hline\hline 
\end{tabular}

\noindent
\underline{Table 1} $\a _{max}^2$ for different $\b $ ($g=0$). $\a _{max}$ 
decreases as we increase $\b $.

\vspace{.1in}

\begin{tabular}{||l|l||}
\hline\hline & \\
$\b $  & $M/evG$\\
\hline\hline &   \\
1.0  &	1.22097 \\
0.5  &	1.19034 \\
0.27 &	1.11625 \\
0.2  &	1.05062	\\	
0.15 &	0.98329 \\
\hline\hline
\end{tabular}

\noindent
\underline{Table 2} $M/evG $ for different $\b $ ($g=0$ and $\a ^2 = 2.5 $).
Mass decreases as we decrease $\b $.

\newpage
\begin{figure}
\vspace{-1in}
\vglue.1in
\makebox{
\epsfxsize=9in
\epsfbox{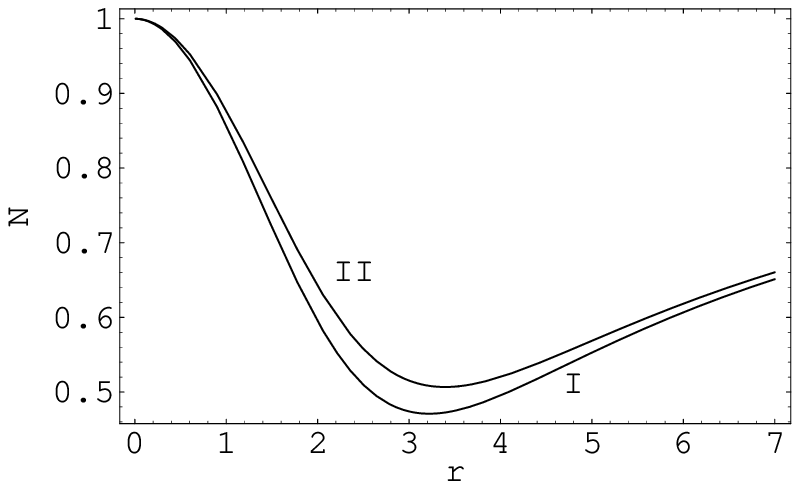}
}
\vspace{-9in}
\caption{Plot for the metric function $N$ as a function of $r$
for $g = 0$, $\a ^2 = 2.5 $ for different values of $\b $. Curve I is
for $\b = 1.0 $ and curve II for $\b = 3.0 $ .
}


\vglue.1in
\makebox{
\epsfxsize=9in
\epsfbox{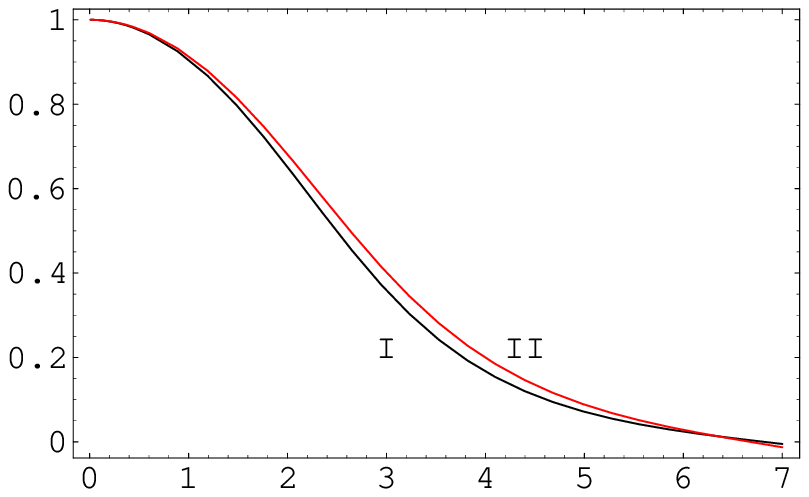}
}
\vspace{-9in}
\caption{Plot for the gauge field $W$ as  a function of $r$
for $g = 0$, $\a ^2 = 2.5 $ for different values of $\b $.  
Curve I for $\b = 1.0 $ and curve II for $\b = 3.0 $ . }

\end{figure}

\newpage

\begin{figure}
\vspace{-9cm}
\vglue.1in
\makebox{
\epsfxsize=9in
\epsfbox{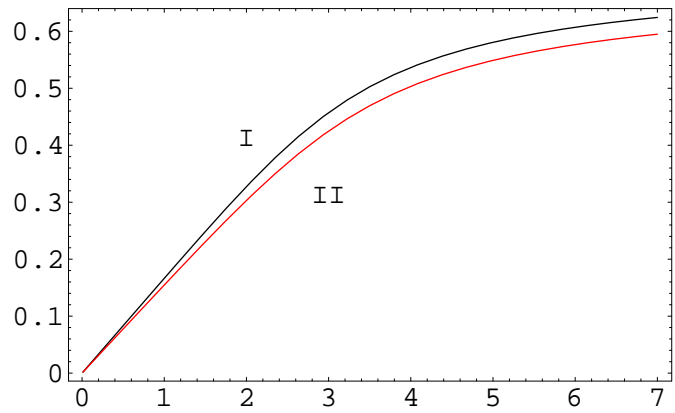}
}
\vspace{-9in}
\caption{Plot for the Higgs field $H$ as a function of $r$
for $g = 0$, $\a ^2 = 2.5 $ for different $\b $.  
Curve I is for $\b = 1.0 $ and curve II for  $\b = 3.0 $ . }
\end{figure}

\end{document}